\documentclass[12pt]{article}

\newcommand{\ket}[1]{|#1\rangle}
\newcommand{\inket}{|\mbox{in}\rangle}
\newcommand{\outket}{|\mbox{out}\rangle}
\newcommand{\bra}[1]{\langle#1|}

\begin{document}

\begin{center}
{\LARGE Counting Marbles With \\ `Accessible' Mass Density:\\ A Reply to Bassi and Ghirardi
\par} \vskip 3em
\lineskip .75em
{\Large Rob Clifton \par} \vskip .75em
{\textit{Departments of Philosophy and History and Philosophy of Science,
University of
Pittsburgh, Pittsburgh, PA 15260 \\ (e-mail: rclifton+@pitt.edu)} \par}
\vskip .75em
{\Large and \par}
{\Large Bradley Monton \par} \vskip .75em
{\textit{Department of Philosophy, 1879 Hall,
Princeton University, Princeton, NJ 08544-1006 \\ (e-mail:
bjmonton@princeton.edu)} \par}
\vskip .75em
\vskip 1.5em%
{\large \today \par}%
\end{center}
\newpage

\section{Review of the Counting Anomaly Debate}

In the previous issue of this journal ([1999]) we argued that, while 
Lewis ([1997]) is correct 
that the enumeration principle fails in dynamical reduction 
theories, one need not follow Lewis in rejecting
these theories. Because the dynamical reduction process itself 
prevents the failure of enumeration from ever becoming manifest, and 
because one can treat the semantics for 
dynamical reduction theories as not adding
anything of ontological import to them, it is reasonable to accept 
these theories notwithstanding Lewis's counting anomaly.  

Lewis had considered a collection of $n$ 
non-interacting marbles in the product state
\begin{equation}
     \ket{\psi}_{\mbox{all}}=  (a\inket_{1} + b\outket_{1})\otimes    (a\inket_{2} +
b\outket_{2})\otimes\cdots\otimes
     (a\inket_{n} + b\outket_{n}),
     \end{equation}
where the state $\inket_{i}$ refers to localization of the $i$th marble 
to inside 
a box (assumed, for simplicity, to be the same box for all marbles), and 
$1>|a|^{2}>>|b|^{2}>0$.   These inequalities are strict 
because dynamical reduction theories generically do 
not eliminate the tails of an object's wavefunction when it 
collapses.  Nevertheless, since $|a|^{2}>>|b|^{2}$, the probability of 
finding any particular marble in the box is very high, and it is 
natural to take each individual marble to be in the box.  Indeed, not 
doing so would leave dynamical reduction theories without a 
complete solution to the measurement problem (see also Albert and 
Loewer [1996]).  On the other hand, 
for sufficiently large $n$, the probability that \emph{all} $n$ marbles will be
found in the box is $|a|^{2n}<<1$, making it highly unlikely
that they will all be
found in the box. Lewis inferred from this that the state $\ket{\psi}_{\mbox{all}}$
`cannot be one in which all $n$ marbles are in the box, since there is almost no chance that if one
looks one will find them there' ([1997], p. 318).  We cashed out 
Lewis's inference by viewing it as a consequence of what we called the 
\emph{fuzzy} eigenstate-eigenvalue link:
\begin{quote}
`Particle $x$ lies in region $R_{x}$ \emph{and} $y$ lies in $R_{y}$ 
\emph{and} $z$ lies in $R_{z}$ \emph{and} $\ldots$'
if and only if the proportion of the total squared
amplitude of $\psi(t, \mathbf{r}_{1},\ldots, \mathbf{r}_{N})$ that 
is associated with points in $R_{x}\times R_{y}\times 
R_{z}\times\cdots$ is greater than or equal to $1-p$.
\end{quote}
The essence of Lewis's counting anomaly, then, is that the 
application of the fuzzy link in state $\ket{\psi}_{\mbox{all}}$ to the collection of marbles 
as a whole conflicts with its application on a marble-by-marble 
basis.  And the only way out of this difficulty is
 to abandon an instance of the logical rule of conjunction 
introduction that Lewis called 
the \emph{enumeration principle}: `if marble 1 is in the box and marble 2 is in
 the box and so on
through marble $n$, then all $n$ marbles are in the box' ([1997], p. 
321).   

Since we were defending this part of Lewis's argument, we were obliged 
to evaluate counter-arguments to it. Thus, 
in section 3 of our
paper, we criticized three arguments against Lewis's claim, the main one being due to
Ghirardi and Bassi ([1999]). 
In their response to our paper, Bassi and Ghirardi ([1999]) (whom we shall henceforth refer 
to simply as `BG') reject our 
criticism, as well as our argument against Lewis that dynamical 
reduction precludes failures of conjunction 
introduction from ever becoming manifest.  Our intention here is to
demonstrate that BG's responses to us do not succeed.

\section{The Counting Anomaly Dies Hard} 

Ghirardi and Bassi's ([1999]) rejection of Lewis's counting anomaly was based, 
in part, upon calculating that, for an object the size of a marble, the 
coefficient 
$a$ occurring in $\ket{\psi}_{\mbox{all}}$ will turn out to be so close in 
absolute square to $1$ that, even if all the mass in 
the actual universe were used to form the marbles, $n$ would not be 
large enough for $|a|^{2n}$ to be appreciably smaller than $1$.  
In our reply ([1999], pp. 17-8), we pointed out that one could equally 
well pose 
the anomaly in terms of counting sufficiently many \emph{microscopic} 
objects, such as 
elementary particles, whose $a$ coefficients would certainly \emph{not} 
be forced by 
dynamical reduction to be exceedingly close to $1$.  Clearly it would be no less strange if arithmetic 
failed for the particles out of which things like marbles are made up!  
Moreover, the virtue of our own response to the 
anomaly (i.e., that it is never manifest) is that it applies in any world governed by dynamical 
reduction, independent of contingent limits on marble or 
particle aggregation in our world.  This leaves us puzzled about why 
BG  
continue to emphasize that the magnitude of the number 
$n$ required to get the counting anomaly off the ground is `unphysically 
large' ([1999], p. 15), and that it is a `peculiar game of putting no limit on the 
number of marbles' ([1999], p. 16).  We echo their sentiment that `we do not like to play with science 
fiction arguments' ([1999], p. 16), but it is not as if one is being 
asked to consider an \emph{actual infinity} of objects.

In any case, Ghirardi and Bassi's ([1999]) main argument against the counting 
anomaly was based upon rewriting the state $\ket{\psi}_{\mbox{all}}$ 
as 
\begin{eqnarray} 
\ket{\psi}_{\mbox{all}} =
a^{n}\inket_{1}\otimes\inket_{2}\cdots\inket_{n}+
a^{n-1}b\outket_{1}\otimes\inket_{2}\cdots\inket_{n}
\nonumber  \\ \label{eq:atbest}
+ a^{n-1}b\inket_{1}\otimes\outket_{2}\cdots\inket_{n}
+\cdots + a^{n-1}b\inket_{1}\otimes\inket_{2}\cdots
\outket_{n} \\
+a^{n-2}b^{2}\outket_{1}\otimes\outket_{2}\cdots\inket_{n}
+\cdots + b^{n}\outket_{1}\otimes\outket_{2}\cdots
\outket_{n}, \nonumber
\end{eqnarray}
and then asserting:
`the precise GRW dynamics will lead in about one millionth of a second to the suppression of the
superposition and to the ``spontaneous reduction'' of the state 
[(\ref{eq:atbest})] to one of its terms'  ([1999], p. 52).
Were such a reduction to occur, conjunction 
introduction would certainly hold for the resultant
state. Therefore, Ghirardi and Bassi concluded that conjunction introduction holds, since a state like (\ref{eq:atbest}) 
`never occurs, its existence being forbidden by the GRW
theory itself' ([1999], p. 52). 
In our paper, however, we argued that Ghirardi and Bassi's characterization of 
the reduction of the state (\ref{eq:atbest}) was 
incorrect and did nothing to remove the anomaly.  
While  BG implicitly concede 
the error ([1999], pp. 10-11), they now claim:
\begin{quote}
\ldots the point we wanted to make is that a rigorous interpretation of states of this type requires a
consideration of the mass density functional \ldots and cannot be based 
on the \emph{fuzzy link}
criterion or other inappropriate criteria. Therefore, the criticism of 
Clifton and Monton ([1999]) to
our previous paper misses the crucial point of our argument just because it does not make
reference to valid criteria for interpreting the wave function within dynamical reduction models
([1999], p. 11).
\end{quote}

We definitely did not see that the crucial point of Ghirardi and 
Bassi's ([1999]) argument against
Lewis was a rejection of fuzzy link-style semantics in favor of the mass density interpretation.
We think we can be forgiven for the oversight given that 
reference to the mass density interpretation in their ([1999]) only 
occurs in 
three places --- two footnotes and a parenthesis --- and, at the 
second occurrence (footnote 3), they state that their arguments 
against Lewis could equally well be made under the mass density 
interpretation (which has the effect of suggesting that their arguments did not 
critically depend upon that 
interpretation).  Moreover, while it may now be the case that the mass 
density approach is `universally accepted' by dynamical reduction 
theorists (BG [1999], p. 7), one can find explicit endorsement in Pearle 
([1997], pp. 150-1) of fuzzy link-type 
semantics applied to quantities such as local charge, spin, energy, 
and momentum (in particular, see equation (11) of Pearle (1997)).  
The important issue, of course, is whether the counting anomaly arises 
under the mass density interpretation (setting aside, for the 
next section, the issue of whether the anomaly can be made manifest).  
We shall argue that either the counting anomaly arises, or some 
other (equally surprising) anomaly will arise. 

We start by reviewing a nice example that Ghirardi, Grassi, and Benatti 
([1995]) use to first
introduce the mass density interpretation. 
Consider two disjoint spherical regions of space $A$ and $B$, each large
enough to contain a macroscopic number $N$ of particles, each of mass 
$m$. Let $\ket{\phi^{A}_{N}}$
denote a state where the $N$ particles are each `well-localized with respect to the characteristic length
($10^{-5}$ cm) of the model and uniformly distributed' in region $A$ 
([1995], p. 16), and 
consider two possible states of the $N$-particle system:
\begin{equation}
\ket{\psi^{+}} = 1/\sqrt{2} (\ket{\phi^{A}_{N}} + 
\ket{\phi^{B}_{N}})\ \  ,\ \ 
\ket{\psi^{\otimes}} =  \ket{\phi^{A}_{N/2}} \otimes \ket{\phi^{B}_{N/2}}. 
\end{equation}
For simplicity, Ghirardi \emph{et al} divide space up into cells with 
dimensions given by the 
characteristic length.  They then consider the expectation of the mass 
operator $M_{i}$ for 
a cell $i$,
\begin{equation}
\mathcal{M}_{i}=\bra{\psi}M_{i}\ket{\psi}.
\end{equation}
It is easy 
to see that both 
$\ket{\psi^{+}}$ 
and $\ket{\psi^{\otimes}}$ give rise to the same mass density 
function for a cell $i$ in region $A$:
 \begin{eqnarray}
\mathcal{M}^{+}_{i} 
& = &  \bra{\psi^{+}}M_{i}\ket{\psi^{+}}\  
\mbox{$\approx$}\   1/2 \bra{\phi^{A}_{N}} M_{i} \ket{\phi^{A}_{N}}\  
 \mbox{$\approx$}\    nm/2, \\
\mathcal{M}^{\otimes}_{i} &
 =   & \bra{\psi^{\otimes}}M_{i}\ket{\psi^{\otimes}}\  
 \mbox{$\approx$}\   
 \bra{\phi^{A}_{N/2}}M_{i}\ket{\phi^{A}_{N/2}}\  
\mbox{$\approx$}\  nm/2,\end{eqnarray}
where $n$ is the number of particles per cell. 
However, Ghirardi \emph{et al} argue that only one of these masses is objective. They consider what happens if
one sends a test mass between regions $A$ and $B$. If the system were in state
$\ket{\psi^{\otimes}}$, gravitational forces would balance and the particle would be undeflected.
On the other hand, if the system were in state $\ket{\psi^{+}}$, 
the test particle would become involved in
the superposition, where in one branch it is deflected toward $A$ and in 
the other toward $B$. Ghirardi \emph{et al} write:
\begin{quote}
Nowhere in the universe is there a density corresponding to the density of the test particle. In a
sense, if one would insist in giving a meaning to the density function, he would be led to conclude
that the particle has been split by the interaction into two pieces of half its density. 
This analysis shows that great attention should be paid in attributing an `objective' status to the
function $\mathcal{M}(\mathbf{r})$  ([1995], p. 17; see also Ghirardi and 
Grassi [1996], p. 365).
\end{quote}
Ghirardi \emph{et al} are then led to propose that the mass in a cell $i$ is `objective' just 
in case $\mathcal{R}_{i}\equiv\mathcal{V}_{i}/\mathcal{M}_{i}^{2}<<1$, where the variance of $M_{i}$ is defined in the usual way by 
\begin{equation}
\mathcal{V}_{i}=\bra{\psi}\left 
[M_{i}-\bra{\psi}M_{i}\ket{\psi}\right ]^{2}\ket{\psi}.
\end{equation}
Only in state $\ket{\psi^{\otimes}}$ is $\mathcal{R}_{i}<< 1$ for a cell in 
$A$; for the state $\ket{\psi^{+}}$, one has $\mathcal{R}_{i}\approx 1$.  

Note 
that $\mathcal{R}_{i}<< 1$ implies that any mass measurement in cell 
$i$ is 
almost certain to yield $mn/2$ as its answer.  In fact, this is what
BG ([1999]) take to be the main motivation for the proposed objective mass 
density criterion; for, in their own introduction to the criterion, 
they reproduce the following passage from Ghirardi 
([1997]): 
\begin{quote}
A property corresponding to a value (or range of values) of a certain 
variable in a given theory is objectively possessed or accessible 
when, according to the predictions of that theory, experiments (or 
physical processes) yielding reliable information about the variable 
would, if performed (or taking place), give an outcome corresponding 
to the claimed value.  Thus the crucial feature characterizing 
accessibility (as far as statements of individual systems is concerned)  is the matching of the claims and the
outcomes of physical processes testing the claims (Ghirardi [1997], p. 227).
\end{quote}
  Surely if these remarks serve to motivate the mass density 
  criterion, they also serve to motivate speaking of a particle (or 
  particles) as being 
located in a region whenever its wavefunction assigns high 
probability to its being detected in that region; that is, they also 
serve to motivate the fuzzy link.  Unfortunately, BG never say why the fuzzy link is 
`inappropriate' and not a `valid' way to understand
reduction theories ([1999], p. 11)\footnote{One might surmise that the fuzzy 
link, which is formulated in terms of configuration, is unacceptable to BG  
because there is no well-behaved relativistic configuration 
operator (e.g., see Malament [1996]). However, it does not matter whether 
one conducts  
discussion of the counting anomaly within a nonrelativistic or 
relativistic setting.  If one wants to insist on the latter, then it is 
trivial to reformulate the 
fuzzy-link in terms of observables that \emph{are} relativistically 
acceptable (local charge, energy, etc.).}. 

It is also unclear, when BG invoke the `physically more 
expressive term' \emph{accessible} ([1999], p. 7), just how seriously we 
are to take talk of mass density as `objective' when the condition 
$\mathcal{R}_{i}<<1$ obtains.  Elsewhere, Ghirardi and Grassi ([1996], 
p. 376) have written, with regard to the term 
`objective', that `both usual
meanings of that term (i.e. ``real'' or ``opposite to subjective'') do not fit with the sense which
emerges for it from our work'. Our response to BG's 
criticism of the 
counting anomaly based on mass density will not in any way 
depend on how these terms are to be understood.  However, it is worth 
reminding the reader of the tentative position that we arrived at in 
our ([1999]): that 
the fundamental ontology of dynamical reduction theories consists only of 
wavefunctions and their evolution, and that a fuzzy link or mass 
density semantics, though essential for connecting the theory 
to our 
everyday experience and conception of the world, need not be seen 
as essential 
for making 
ontological sense of such theories.

We return, now, to the counting anomaly.  In order for mass density 
to be relevant to the issue of how many particles are in the 
box, we need a way to connect mass density talk to position talk.
 Ghirardi \emph{et al} already mention the obvious way of doing this: 
 `The reality of a massive macro-object in front of us corresponds to the fact that in the region it
occupies there is the objective mass density which characterizes it'  
([1995], p. 33).
And this is BG's understanding of things, when they assert:
\begin{quote}
\ldots the mass is objective precisely in the regions 
where the various marbles are\ldots The marbles can therefore be 
claimed to be all within the box, and the total mass within the box is 
actually the one corresponding to all of them being in the box 
([1999], p. 9).
\end{quote}
It is the second sentence with which we disagree.  Consider, first, 
 a single
marble in state
     $a\inket + b\outket$, where $|a| \gg |b|$.
If the marble's total mass, i.e., mass density function integrated 
over all space, is $m$, then the total accessible mass in the box will only 
be $|a|^{2} m$, and this can of course change with time. If one thought that an essential property of 
a marble is that its objective (rest) mass is constant, one 
might resist locating the marble entirely in the box.  We are 
not inclined to resist that ourselves, since the task, as we 
see it, is to supply a servicable mapping between the theory and our 
language about marbles.  However, the plot thickens when one considers 
an $n$-marble system in state $\ket{\psi}_{\mbox{all}}$.  In that case, 
Lewis ([1997], appendix A) 
provided a detailed analysis, after which he arrived at the same conclusion as BG (!):
\begin{quote}
There are $n$ regions of objective mass $|a|^{2}m$ in the box, one 
corresponding to each marble, and no regions outside the box; 
consequently the mass distribution is such that all $n$ marbles are in 
the box (1997, p. 327).
\end{quote}  
Unfortunately, we fail to see how Lewis's conclusion here follows.  Since the total mass 
in the box is $(|a|^{2}m)\times n$,  $n$ can be chosen sufficiently large 
so that 
total 
does not exceed, for example, $m(n-1000)$. Since one could well believe that 
it is inappropriate to assert 
that all $n$ marbles are 
in the box unless roughly $mn$ mass is objective therein, conjunction 
introduction need not be satisfied.  And it is little help to
respond that, for $n$ sufficiently large, the number $m(n-1000)$ \emph{is} 
roughly the same as
 $mn$.  For one could still be puzzled as to why it would not 
then be 
equally legitimate to assert that $n-1000$ marbles are in the box 
as well.  

It should be emphasized that even though Lewis claimed not to find 
any failure of
conjunction introduction under the mass density 
interpretation, he maintained that it was unacceptable to assert that 
all marbles are in the box, on the grounds that there is still a 
vanishing probability in state $\ket{\psi}_{\mbox{all}}$ of finding them all there (1997, 
p. 320).  Though BG are dismissive of Lewis on this point ([1999], note 2), we entirely agree 
with him (and gave supporting arguments 
for his position in our [1999], 
pp. 21-2).  Note also that if one wants to maintain that all the mass of 
the marbles is objectively in the box, together with the fact that the 
probability of finding them all there is vanishingly small, then one is 
committed to a radical breach between mass and location talk. Thus we can 
pose BG a trilemma for the 
mass density interpretation: either conjunction introduction fails, 
mass talk must be divorced from position talk, or the intuitive
connection between either of these kinds of talk and a system's 
dispositions (cf. `accessibility') must be severed. 

\section{Suppressing the Counting Anomaly}

In our ([1999]), we defended the fuzzy link in the face of Lewis's counting 
anomaly by showing that a violation of conjunction
introduction could never be made manifest.  Notwithstanding BG's 
critique, we stand by our
arguments.  Indeed, we believe that they apply equal well in defence of the mass 
density criterion! 

Our escape from the counting anomaly was based on Lewis's failure to 
explicitly model, or `operationalize', the counting 
process itself within dynamical 
reduction theories.  BG begin their critique by rejecting this line of thinking out of hand:
\begin{quote}
The unappropriateness of such a point of view has already been 
stressed in section 3.  Within GRW's theory no \emph{measurement} process ever occurs, the only physical processes
being interactions among physical systems governed by universal laws 
([1999], p. 17).
\end{quote}
Thus, it is worth being clear 
that nothing in our analysis turned on privileging 
(so-called) \emph{measurement} interactions.
  
In that analysis, we considered 
what would happen when devices $M_{i}$, each designed to register whether 
or not 
marble $i$
is in the box, are coupled to the marbles in 
state $\ket{\psi}_{\mbox{all}}$; and when a  system $M$, designed to register 
the \emph{total} number of particles in the 
box,  is coupled to the total $n$-marble system.  Just after interaction with all the registers, the state 
would be (using an obvious notation):
\begin{eqnarray} \label{eq:sixteen}
\ket{\psi}_{\mbox{count}} & =  &
a^{n}\ket{\phi}_{\mbox{out}_{0}}\ket{\mbox{`}O=n\mbox{'}}_{M}+
a^{n-1}b\ket{\phi}_{\mbox{out}_{1}}\ket{\mbox{`}O=n-1\mbox{'}}_{M} \\
\nonumber & &
+\cdots+
b^{n}\ket{\phi}_{\mbox{out}_{n}}\ket{\mbox{`}O=0\mbox{'}}_{M}, \end{eqnarray}
where
\begin{eqnarray} \nonumber
     \ket{\phi}_{\mbox{out}_{0}} & = &
\inket_{1}\ket{\mbox{`in'}}_{M_{1}}\inket_{2}\ket{\mbox{`in'}}_{M_{2}}\cdots
     \inket_{n}\ket{\mbox{`in'}}_{M_{n}}  \\ \nonumber
          \ket{\phi}_{\mbox{out}_{1}} & = &
     \ \ \outket_{1}\ket{\mbox{`out'}}_{M_{1}}
     \inket_{1}\ket{\mbox{`in'}}_{M_{2}}\cdots
     \inket_{n}\ket{\mbox{`in'}}_{M_{n}} \\ \nonumber
          &  &
     +\inket_{1}\ket{\mbox{`in'}}_{M_{1}}
     \outket_{1}\ket{\mbox{`out'}}_{M_{2}}\cdots
     \inket_{n}\ket{\mbox{`in'}}_{M_{n}} \\ \nonumber
     & & \vdots  \\ \nonumber
          &  &
     +\inket_{1}\ket{\mbox{`in'}}_{M_{1}}
     \inket_{1}\ket{\mbox{`in'}}_{M_{2}}\cdots
     \outket_{n}\ket{\mbox{`out'}}_{M_{n}} \\ \nonumber
     \vdots & & \\ \nonumber
     \ket{\phi}_{\mbox{out}_{n}} & = &
     \outket_{1}\ket{\mbox{`out'}}_{M_{1}}
     \outket_{2}\ket{\mbox{`out'}}_{M_{2}}\cdots
     \outket_{n}\ket{\mbox{`out'}}_{M_{n}},
     \end{eqnarray}
     We then argued that dynamical reduction will rapidly reduce the 
     entangled state $\ket{\psi}_{\mbox{count}}$ to one in which 
     the individual recordings of the $n$ 
     registers always agree with whatever total $M$ records.  The argument 
     depends only the fact that, \emph{unlike} 
     $\ket{\psi}_{\mbox{all}}$,
     $\ket{\psi}_{\mbox{count}}$ is an \emph{entangled} state involving a 
     macroscopic number of particles.  Moreover, one need not assume that $M$, or any of the $M_{i}$, are themselves 
     macroscopic (cf. our [1999], notes 6,7).  Of course, in order for the marble count(s) to be 
     \emph{known} by the `gross creatures that we are' (to echo 
     a phrase of John Bell's), these devices would have to be coupled to 
     macroscopic devices.  But the suppression of the counting anomaly will 
     have occurred long before information about the marbles ever 
     enters anyone's consciousness --- it is the purely physical 
     outcome of purely physical interactions and correlations. 
     
     Criticizing our discussion of the various possible collapse scenarios 
     that could follow interaction with the registers, BG ask rhetorically:
     \begin{quote}
     The marbles are macroscopic systems 
     and their masses may very well be comparable or even larger than 
     those of the pointers both of the apparatuses $M_{i}$ and of $M$.
      Why the GRW dynamics has to be suspended 
     up to the time in which the pointer of $M$ is
localized, and subsequently (in the case that $O\not=n$) up to the 
moment in which all pointers of the $M_{i}$'s are localized, is really 
a big mystery ([1999], pp. 20-1).
\end{quote}
  In our exposition, we were simply 
dividing the collapse process into different stages for ease of 
exposition, not trying to give a real-time description of 
the dynamics.   Certainly we were aware that the marbles themselves will be almost 
continually subject to GRW collapses before, during, and after the interaction 
with the registers, causing  marbles to jump in and out of the box 
almost all the time (provided enough marbles are considered).  As BG 
say, `The situation\ldots changes every millionth of a second' 
([1999], p. 16).  
However, we fail to see 
why this means we:
\begin{quote}
\ldots must first of all assume that it is 
possible to perform the incredible task of testing the positions of 
all particles and the total mass within the box in much less than a 
millionth of a second ([1999], p. 16).
\end{quote}
For, as we argued ([1999], pp. 18-9), at \emph{every} moment prior to 
the entangling 
interaction with $M$, there will always be some conjunctive 
assertion about the marbles' locations relative to the box with 
respect to which conjunction 
introduction fails.  The point of introducing $M$ was not so it can 
provide a faithful recording of the total marble count \emph{prior} to its 
interaction with the marbles, but to demonstrate that the \emph{most 
current} marble count 
will always agree with the \emph{most current} information contained in the $M_{i}$ 
registers.   
   
BG 
summarize our response to Lewis with the phrase: `the violation of the counting 
rule is still there, but it cannot be revealed' ([1999], p. 20).  This 
is not correct: the 
marbles' violation of the enumeration principle will itself 
disappear, along with any chance of \emph{manifesting} such a violation, a 
split second after their interaction with $M$ (i.e., as soon as 
 $\ket{\psi}_{\mbox{count}}$ collapses).  
Maintaining their resistance to this conclusion, BG finish their 
critique by apparently arguing that a failure of conjunction introduction in 
the marbles system
 could, after all, be
made manifest:
\begin{quote}
Since $M$'s reading is given by its pointer location \ldots a consistent use of the GRW dynamics
implies that \emph{it cannot} be perfectly localized: it will be affected by the `tails problem' just as
all macroscopic systems. Thus, if one follows the authors in their strange argument, one should
conclude that (with a small probability) the state can be the one in which $n$ marbles are in the
state $\inket$, $n$ apparatuses $M_{i}$ register $\ket{\mbox{`in'}}$, but the pointer of the
apparatus $M$ can be reduced on a state corresponding to its pointing at 
$k \not= n$! ([1999], p. 21)
\end{quote}
In our analysis, we ignored the subsequent development of tails in 
the wavefunction of  $M$'s pointer after its interaction with the 
marbles, and treated that interaction as `perfect' in the sense 
that, if the marbles had been in an eigenstate of a definite number of 
particles, $j$, in the box, $M$ would register $O=j$.  Dropping these idealizations 
would simply mean that $M$ has some 
small probability of registering an incorrect result for the total 
marble count.  But that kind of measurement error is strictly irrelevant to
manifesting a
failure of enumeration.  Indeed, our point was that even if one 
assumes that the registers operate as perfectly as could be, they will 
not display any counting anomaly.  As we put it in the paper (in 
relation to an observer playing the role of $M$): `the only way to 
arrange things so that our observer \emph{could} falsify the 
enumeration principle would be to suppose that she was never a 
competent enumerator to begin with!' ([1999], p. 31). 

\begin{center}
 \textbf{References}
 \end{center}

\noindent Albert, D. and Loewer, B. [1996]:
`Tails of Schr\"{o}dinger's Cat',  in R.
Clifton (\emph{ed}), \emph{Perspectives on Quantum Reality}, Dordrecht: Kluwer, pp. 81-91.\vspace{.1in}

\noindent Bassi, A., and G. C. Ghirardi [1999]: `More About Dynamical Reduction and the
Enumeration Principle', \emph{The British Journal for Philosophy of 
Science}, December issue. \vspace{.1in}

\noindent Clifton, R. and B. Monton [1999]: `Losing Your Marbles in Wavefunction Collapse
Theories',  \emph{The British Journal for Philosophy of Science}, 
December issue. \vspace{.1in}

\noindent Ghirardi, G. C.  [1997]: `Macroscopic Reality and the 
Dynamical Reduction Program', in
M. L. Dalla Chiara \emph{et al} (eds), \emph{Structures and Norms in 
Science},
Dordrecht: Kluwer, pp. 221-240. \vspace{.1in}

\noindent Ghirardi, G. C. and A. Bassi [1999]:
`Do Dynamical Reduction Models Imply That Arithmetic Does Not Apply to Ordinary
Macroscopic Objects?',
\emph{The British Journal for Philosophy of Science}, \textbf{50}, 
pp. 705-720.
\vspace{.1in}

\noindent Ghirardi, G. C. and R. Grassi [1996]: `Bohm's Theory versus Dynamical Reduction', in
J. T. Cushing \emph{et al} (eds), \emph{Bohmian Mechanics and Quantum Theory: 
An Appraisal},
Dordrecht: Kluwer, pp. 353-377. \vspace{.1in}

\noindent Ghirardi, G. C., R. Grassi, and F. Benatti [1995]: `Describing the Macroscopic World:
Closing the Circle within the Dynamical Reduction Program', \emph{Foundations of Physics},
\textbf{25}, pp. 5-38. \vspace{.1in}

\noindent Lewis, P.  [1997]:
`Quantum Mechanics, Orthogonality, and Counting',
\emph{The British Journal for Philosophy of Science}, \textbf{48},
pp. 313-328.\vspace{.1in}

\noindent Malament, D. [1996]: `In Defense of Dogma: Why There Cannot be a 
 Relativistic Quantum Mechanics of (Localizable) Particles', in R. Clifton 
 (\emph{ed}), \emph{Perspectives on Quantum
 Reality}, Boston: Kluwer, pp. 1-10.\vspace{.1in}

\noindent Pearle, P. [1997]:
`Tales and Tails and Stuff and Nonsense',  in R. S. Cohen, M. Horne, 
and J. Stachel (\emph{eds}), \emph{Experimental Metaphysics}, Dordrecht: 
Kluwer, pp. 143-56.\vspace{.1in}

\end{document}